\documentclass{elsart}

\usepackage{graphicx,amssymb,amsmath}
\usepackage{cite}
\usepackage{bm}

\begin{document}
\begin{frontmatter}
\title{Charge independence studied in $NN\to d\pi$ reactions}
\author[a,b]{H.~Machner}\corauth[cor]{Corresponding author}\ead{h.machner@fz-juelich.de},
\author[c]{J. A. Niskanen},
\address[a]{Institut f\"{u}r Kernphysik, Forschungszentrum J\"{u}lich, 52425 J\"{u}lich, Germany}
\address[b]{Fachbereich Physik, Universit\"{a}t Duisburg-Essen, Lotharstr. 1,
 47048 Duisburg, Germany}
\address[c]{Department of Physical Sciences, University of Helsinki, Finland}
\begin{abstract}
We review to what extent charge independence breaking (\emph{CIB})
in isospin related reactions of the type $NN\to d\pi$ can or could
be seen in existing data. In doing this we present fits to global
and threshold cross sections including most recent data. Applying
these we point out the probable impossibility to make model
independent predictions even based solely on the different threshold
energies and the well known $pp\rightarrow d\pi^+$. A possible
discrepancy is seen between the $np\rightarrow d\pi^0$ and
$pp\rightarrow d\pi^+$ data, which may require invoking explicitly
isospin symmetry breaking interactions.

\end{abstract}
\end{frontmatter}
\section{Introduction}
It is well known that the isospin invariance as introduced
originally by Heisenberg is only an approximate symmetry even in the
case of strong interactions. While in nuclei it is relatively
strongly broken, it is, however, very good on the level of nucleons,
where the mirroring of neutrons and protons is related to charge
symmetry (\emph{CS}), a more accurate symmetry than charge
independence (\emph{CI}). In the framework of the meson theory of
nuclear forces a classification of different isospin structures
 is given in Ref. \cite{Henley_Miller79}.
After Weinberg's ground breaking work on the quark masses
\cite{Weinberg77} it became clear that in addition to the Coulomb
force, the difference between the light quark masses breaks isospin
symmetry. This means that in QCD under exchange of the $u$ and $d$
quarks there will be physically observable changes. Therefore,
 measurements of the violation of the
isospin symmetry might allow the deduction of this mass difference.
Recent reviews related mainly to this charge symmetry breaking
(\emph{CSB}) are given in Refs. \cite{Miller90,Miller95}. For a
recent general comprehensive review on meson production we refer to
\cite{Machner99} and \cite{Hanhart04}.

Isospin symmetry breaking effects are small and few dedicated
experiments have been performed to study them in inelasticities.
Nevertheless, in this article we attempt to draw together what is
known about the relation between the two reactions
\begin{equation}\label{equ:np}
np \to\pi^0d
\end{equation}
and
\begin{equation}\label{equ:pp}
pp \to\pi^+d
\end{equation}
paying special attention to their differences. It is readily clear
that these are {\em not} related by the charge symmetry operation (a
reflection of the isospin in the $z$ or 0 direction), but by charge
independence, a wider symmetry of arbitrary rotations in the isospin
space.

As pointed out by Yang \cite{Yang52} already in 1952 isotopic spin
invariance requires that the cross sections must be related by
\begin{equation}\label{equ:iso_ratio}
\frac{d\sigma(pp \to\pi^+d)}{d\sigma(np \to\pi^0d)}=2.
\end{equation}
This can be shown by expressing the initial and final states in
terms of eigenstates of the total isospin (spin summation implied)
\begin{equation}\label{equ:Ampli_ratio}
\frac{d\sigma(pp \to\pi^+d)}{d\sigma(np \to\pi^0d)}=\frac{|\langle
1,1|S|1,1\rangle |^2}{| 1/\sqrt{2}\, \langle 1,0|S|1,0\rangle
+1/\sqrt{2}\, \langle 1,0|S|0,0\rangle |^2}
\end{equation}
where $S$ is the scattering matrix. If the isospin is conserved as
required by charge symmetry, the second term in the denominator must
vanish. Also, if the strong forces are independent of the third
component of isospin, i.e. charge independent, then
\begin{equation}
\langle 1,1|S|1,1\rangle =\langle 1,0|S|1,0\rangle ,
\label{chargeindep}
\end{equation}
and the amplitudes in the numerator and denominator cancel leaving
the factor of two.

\emph{CSB} requires typically a vector operator (tensor of rank one)
in the isospin space as explicitly seen in the classification of two
nucleon forces in Ref. \cite{Henley_Miller79}, i.e. the isospin
structures $(\bm{\tau}_1 +\bm{\tau}_2)_0$ (class III) or
$(\bm{\tau}_1-\bm{\tau}_2)_0$ or $(\bm{\tau}_1\times\bm{\tau}_2)_0$
(class IV). Here the latter operators necessarily change the isospin
and thus also in the $np\to d\pi^0$ reactions (and only there) can
mix isospin zero initial state in the production amplitudes. This
causes an asymmetry of the cross section about $90^\circ$ in the
centre of mass system. Otherwise the cross section would be
symmetric \cite{Barshay_Temmer64}. A measurement of the charge
symmetry breaking asymmetry in the meson production reaction $np \to
\pi^0 d$ close to threshold has been published recently
\cite{Opper03}.

Charge independence is a broader symmetry encompassing any
combination of charges, {\it i.e.} a symmetry with respect to
arbitrary rotations in the isospin space. Isotensor class II forces
of Ref. \cite{Henley_Miller79} $(3\tau_{10}\tau_{20} -
\bm{\tau}_1\cdot\bm{\tau}_2)$, while respecting charge symmetry, do
violate charge independence in the two nucleon case, and similarly
again Eq. (\ref{chargeindep}) need not be valid in meson production
either. A deviation from the factor of two would be a sign and a
measure of violation of isotopic invariance. It should be noted that
in the present case and in $NN$ scattering the isotensor operator
cannot change the isospin, which has to be unity in both the initial
and final state. Therefore, the spin-spatial symmetries would be the
same as in the isospin symmetric case, contrary to charge symmetry
breaking.

Unfortunately even less is available for charge independence
breaking (\emph{CIB}) than \emph{CSB} in meson production, although,
in principle, this should be significantly larger than \emph{CSB}, a
few \% vs. a few thousandths. To our knowledge, only the
unsuccessful attempt \cite{Hollas81} to see \emph{CSB} reports
possible marginal \emph{CIB} in the angular dependence at 795 MeV.
The experimental difficulties here stem from two origins: Firstly,
there is not a "null experiment" nor such a clear forward-backward
asymmetry signal as in \emph{CSB}. Secondly, these measurements
require clearly a comparison of separate experiments, one with
neutron beams, prone to meet normalisation problems.

However, further motivation to study isospin symmetry breaking for
the above reactions (\ref{equ:np}) and (\ref{equ:pp}) is given by
the recently measured, somewhat related more complicated reactions
\begin{equation}
pd\to \pi^0\ ^3{\rm He} \label{eqn:pi0}
\end{equation}
and
\begin{equation}
pd\to \pi^+\ ^3{\rm H} \, . \label{eqn:pi+}
\end{equation}
These reactions have the advantage that they can be measured
simultaneously and thus their ratio does not suffer from
normalisation uncertainties. Indicative of isospin symmetry
breaking, the excitation function of the ratio between the
differential cross sections for maximal momentum transfer to the
pion showed deviation from the isospin ratio two when passing the
$pd\to \eta ^3{\rm He}$ threshold \cite{Abdel-Bary03}. Full angular
distributions for both reactions at lower energies were reported in
Refs. \cite{Betigeri01} and \cite{Abdel-Samad03}. There the angular
distributions also show a slight but systematic difference. However,
the origin of this difference could be different nuclear wave
functions (possibly due to the Coulomb force) of $^3$H and $^3$He
rather than \emph{CIB} production mechanisms. It would be
interesting to see if and how isospin symmetry breaking would
persist in the more basic and theoretically cleaner reactions
(\ref{equ:np}) and (\ref{equ:pp}) without this distortion.

The basic theoretical arguments for handling the data are presented
in the next Section and then the results of the analyses in Section
3.

\section{Coulomb corrections and kinematic considerations}
Differential as well as total cross sections of the two reactions
obviously differ because of the mass differences of the nucleons in
the entrance channel and of the pions in the exit channel, and also
due to the presence of the Coulomb force in the reaction
(\ref{eqn:pi+}). Furthermore, charge dependent forces may be
involved in the reactions. Typically these would arise from charged
vs. neutral meson mass differences in meson exchanges. Charge
asymmetry, in turn, can get contributions from nucleon mass
differences, meson mixing \cite{Niskanen99}, magnetic interactions
and explicitly isospin symmetry violating pion-nucleon rescattering
\cite{Kolck00}.

\begin{figure}[t]
\centering
\includegraphics[width=8 cm]{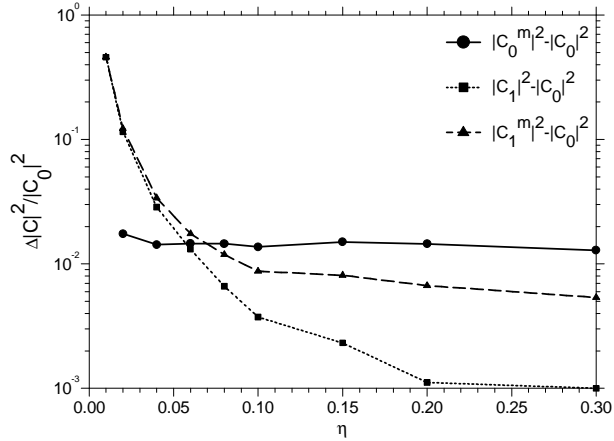}
\caption{Relative differences between the Gamow factors (see Ref.
\cite{Niskanen97}).} \label{fig:Coulomb_Corrections}
\end{figure}
In comparisons of the reactions (\ref{equ:np}) and (\ref{equ:pp}) we
shall first study the second effect. A common approach to correct
for the Coulomb effect is to apply the $s$-wave Gamow penetration
factor
\begin{equation}
 |C_0(\xi)|^2 = \frac{2\pi\xi}{e^{2\pi\xi}-1}\quad {\rm with}\quad
\xi=\frac{\alpha m_{\rm red}c^2}{\hbar cq}\approx\frac{0.0068}{\eta}
\label{gamow}
\end{equation}
to the cross sections. Here $q$ is the centre-of-mass momentum of
the pion and $\eta = q/m_{\pi^+}$.\footnote{One should note that the
final state dependence of the partial wave amplitudes is through the
pion momentum $q$ and that in $\eta$ the charged pion mass acts only
as a unit and should be kept the same in both reactions
(\ref{equ:np}) and (\ref{equ:pp}). Use of different scaling masses
would cause an artificial effect based on the choice of units.
Therefore, the use of "natural"\ mass $m_{\pi^0}$ in reaction
(\ref{equ:np}) is an unfortunate choice.}

Niskanen and Vestama \cite{Niskanen97} have calculated more
generally such factors under various assumptions: for a point-like
charge distribution with pion angular momentum $l$ ($|C_l|^2$) and
for an extended charge distribution ($|C_l^{\rm e}|^2$) applying the
Reid potential \cite{Reid68} to obtain the deuteron wave function.
It was seen that the extended and point sources gave very similar
corrections so that the use of an extended source is not really
warranted with the experimental accuracy expected in near future. In
Fig.~\ref{fig:Coulomb_Corrections} we show the relative difference
of the Gamow factors for $p$-waves with extended charge $|C_1^{\rm
e}|^2$ and $s$-waves $|C_0^{\rm e}|^2$ relative to $|C_0|^2$ as a
function of the dimensionless centre of mass pion momentum
$\eta=p_\pi/m_{\pi^+}$. The first difference is large only in a
range for very small momenta $\eta \leq 0.05$, where $p$-wave cross
section is negligibly small, and then the difference between the two
factors decreases below $2\%$. This is the order of magnitude of
precision that can possibly be obtained.

Justified by these results in removing the Coulomb, for the sake of
simplicity, we will make use of only $|C_0^{\rm e}|^2\approx
|C_1^{\rm e}|^2\approx 1.008\, |C_0|^2$, since a separation into
different partial waves is far from straightforward
\cite{Niskanen97,Machner98a}. As can be seen from
Fig.~\ref{fig:Coulomb_Corrections} in relative angular dependence
this average factor cancels off to a good approximation for extended
sources, since $|C_1^{\rm e}|^2$ and $|C_0^{\rm e}|^2$ are within
0.5\% from each other. The similar factor $|C_0|^2$ is also applied
for the $pp$ initial state giving a normalisation effect of 1-2\%.

\begin{figure}[tb]
\begin{center}
\includegraphics[width=8 cm]{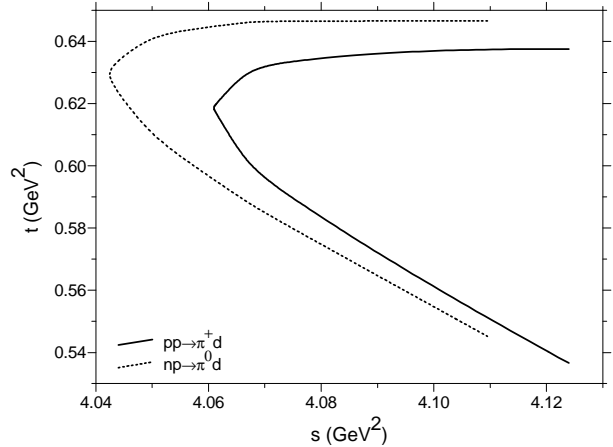}
\caption{Kinematical limits of the two reactions. $t_{max}$ and
$t_{min}$ are shown as function of $s$.} \label{fig:s_t}
\end{center}
\end{figure}

We proceed now to study the effect of kinematical differences due to
different masses. The differential cross section $d\sigma/ dt$
depends on the Mandelstam variables $s$ and $t$. Here we use $t$ as
the four momentum transfer squared from the projectile to the pion.
In Fig.~\ref{fig:s_t} the kinematical limits, i.~e. maximal and
minimal values of $t$ as functions of $s$ are given for the two
reactions (\ref{equ:np}) and (\ref{equ:pp}).

\begin{figure}[tb]
\begin{center}
\includegraphics[width=8 cm]{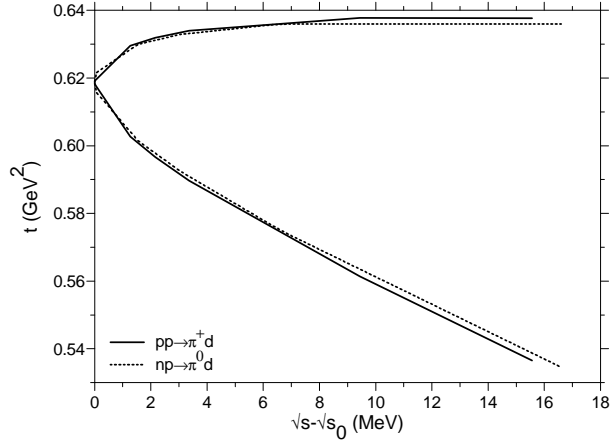}
\caption{Kinematical limits as function of the excess energy
$\epsilon=\sqrt{s}-\sqrt{s_0}$. The $t$-scale is shifted for
reaction (\ref{equ:np}) by $-0.01065$ GeV$^2$. }
\label{fig:sqrt_s_t}
\end{center}
\end{figure}

\begin{figure}[tb]
\begin{center}
\includegraphics[width=8 cm]{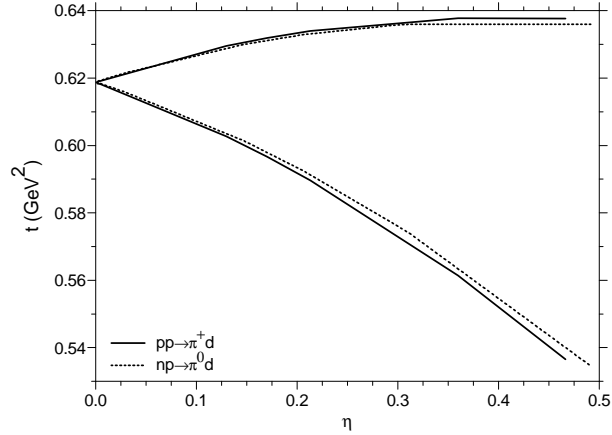}
\caption{Same as Fig.~\ref{fig:sqrt_s_t} but as function of the
dimensionless pion momentum $\eta = q/m_{\pi^+}$.} \label{fig:eta_t}
\end{center}
\end{figure}

In addition to the different threshold energies $\sqrt{s_0}$  also
for given CMS excess energies $\epsilon = \sqrt{s} - \sqrt{s_0}$ the
$t$ values are shifted to a good approximation by a constant 0.01065
GeV$^2$ as can be seen from Fig. \ref{fig:sqrt_s_t}, where the
limits for the reaction (\ref{equ:np}) have been brought on top of
the reaction (\ref{equ:pp}). As a function of the final state
momentum the rounded bottom of the lopsided parabola gets a V shape
in Fig.~\ref{fig:eta_t}.


One may well argue that a large part of the difference between the
two reactions comes, in addition to the Coulomb effects already
described, simply from the different thresholds. This means that for
a given final state momentum (the decisive dependence close to
thresholds) the incident kinetic energy is slightly different. This
gives a trivial effect in the phase space factor $P$ (with an
obvious notation for the energies and momenta $p^*$ in the
centre-of-mass system, $N=p$ or $n$, and including also the spin
factor 1/4)
\begin{eqnarray}\label{eqn:matrix}
\frac{{d\sigma (\theta )}}{{d\Omega }}\, &=&
\underbrace{\,\frac{1}{4(2\pi)^2\hbar^4}\frac{{p_\pi^{*} }}{{p_N^{*}
}}\frac{{E_N E_p E_d E_\pi }}{s} }\underbrace{\sum_{\mu
SM}\mid\langle\psi_d^\mu\mid H^\pi\mid\phi^{SM}\rangle\mid^2}
 \\
& \equiv & \,\,\,\,\,\,\,\,\,\,\,\,
\,\,\,\,\,\,\,\,\,\,\,\,\,\,\,\,\,P_{Np}
\,\,\,\,\,\,\,\,\,\,\,\,\,\,\,\,\,\,\,\,\,\,\,\,\,\,\,\,\,\,\,\,\,\,\,\,\,\,\,\,\,\,\,
\,\,\,\,\,\,\,\,R_{Np}
\end{eqnarray}
present in cross sections, but it influences also the transition
matrices in $R$ themselves. In particular, due to the lower
threshold as a function of the final state momentum the amplitudes
for $np \to\pi^0d$ should lag behind $pp \to\pi^+d$ in reaching and
passing e.g. the $\Delta$ resonance. (One should note that the mass
difference of the nucleons does not directly affect the proximity of
the $\Delta$, since similar mass differences would be present in
that, too.)

In Ref. \cite{Niskanen97} changes in the total and differential
cross section were expressed directly by differences in both the
phase space and spin amplitudes as
\begin{equation}
 \delta\sigma \equiv 2\sigma(np)-\sigma(pp)
=\frac{P_{np}+P_{pp}}{2}(R_{np}-R_{pp})
+(P_{np}-P_{pp})\frac{R_{np}+R_{pp}}{2}\, .
\end{equation}
Results for the relative (integrated) differences in $P$ and $R$
were given, for the latter by a model calculation.

One might choose another path aiming at a model independent
prediction for the difference based on existing data. First dividing
the trivial phase space factor off directly from the empirical data
as well as the Gamow factors (\ref{gamow}) one can study the energy
dependence of the more interesting $R$ for equal pion momenta in the
two reactions. Then one would get from
\begin{equation}
\delta R \equiv R_{np}-R_{pp} = \frac{\mathrm{d} R_{pp}}{\mathrm{d}
E_i}\, ( E_i(np) - E_i(pp))  \label{equ:prediction}
\end{equation}
a prediction for the dynamical difference between the reactions
(\ref{equ:np}) and (\ref{equ:pp}) from data on only the latter.
However, if the quantity $R_{pp}$
 is given as a function of $\eta$
like usual for the cross section, a singular $d\eta /dE_i$ arises.
Consequently this product close to threshold is too large for the
first order estimate to be useful.

It would now be tempting to think that deviations from this
prediction would, in principle, be a direct indication of charge
dependent forces. However, this expectation is based on the
kinematic relation between $E_i$ and $\eta$ used to connect the two
different reactions. It still leaves untouched the possible
dynamical difference in the initial states due to the change of the
energy but without any change in the final momentum. This will be
seen later in Fig. \ref{fig:prediction}.


\section{Data Analysis}
\begin{figure}[tb]
\begin{center}
\includegraphics[width=8 cm]{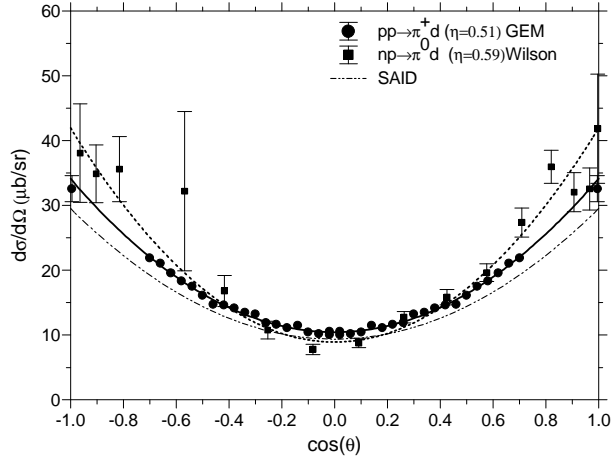}
\caption{Comparison of differential cross sections for
reaction(\ref{equ:np}) shown as squares and for reaction
(\ref{equ:pp}) shown as dots. The solid and long dashed curves are
parameterisations discussed in the text. The short dashed curve is
the SAID-fit \cite{SAID} at $\eta = 0.51$.} \label{fig:Comp_850}
\end{center}
\end{figure}

We are now ready to compare data for the two reactions of interest.
Unfortunately there are no differential cross sections for the same
$\eta$ or $\epsilon$ and the same cm emission angle. Nevertheless,
in order to stress this shortage we compare two angular
distributions. at reasonably similar energies. Fig.
\ref{fig:Comp_850} shows the centre-of-mass differential cross
sections from the data of Wilson et al. for the reaction
(\ref{equ:np}) \cite{Wilson71a} together with those for the reaction
(\ref{equ:pp}) from the GEM collaboration at COSY \cite{Betigeri01a}
(reflected about $\cos\theta=0$).

The differential cross sections can be expanded in terms of Legendre
polynomials $P_l(\cos\theta)$ \cite{Niskanen80}
\begin{eqnarray}\label{equ:Leg_expansion}
4\pi\frac{d\sigma(\eta,\cos\theta)}{d\Omega} &=& A_0(\eta)
P_0(\cos\theta) + A_2(\eta) P_2(\cos\theta) \\
&=&\sigma(\eta) \left[1+a(\eta) P_2\left(\cos\theta\right) \right],
\label{sigexpansion}
\end{eqnarray}
with the indicated normalisation
\begin{equation}
\sigma (\eta)= A_0(\eta)
\end{equation}
and the anisotropy
\begin{equation}
a(\eta) = A_2(\eta)/ A_0(\eta)\, .
\end{equation}
Eq. (\ref{equ:Leg_expansion}) is valid for both the reaction $pp
\to\pi^+d$ and for reaction $np \to\pi^0d$ with the same
coefficients, if charge independence is obeyed (apart from the
normalisation factor 2). As an example this functional dependence is
fitted to the data in Fig. \ref{fig:Comp_850}. The uncertainties of
the fit parameters are compiled in Table
\ref{tab:fit_uncertainties}.
\begin{table}[tb]
\centering \caption{Uncertainty of the fitted parameters Eq.
(\ref{equ:Leg_expansion}) to the data shown in Fig.
\ref{fig:Comp_850}.}\label{tab:fit_uncertainties}
\begin{tabular}{lccccc}
\hline
reaction & data  & $\eta$ & $\Delta A_0/A_0$ & $\Delta A_2/A_2$  & $\Delta a/a$ \\
\hline
$np\to\pi^0d$ & \cite{Wilson71a} & 0.59 & 0.027 & 0.064 & 0.048 \\
$pp\to\pi^+d$ & \cite{Betigeri01a} & 0.51 & 0.0076 & 0.023 & 0.018 \\
\hline
\end{tabular}
\end{table}

The error in the charged total cross section would be small enough
to see isospin breaking effects, while that for the $np$ reaction is
of the same order as those effects computed in Ref.
\cite{Niskanen97}. Furthermore, one should note that the $np$ data
are not absolute in their normalisation. The error in $A_2$ is still
significantly larger. However, interestingly the relative error of
the anisotropy in the form (\ref{sigexpansion}) is smaller than what
would be obtained from the form (\ref{equ:Leg_expansion}).
Apparently this reflects the fact that normalisation uncertainties
present in $A_2(\eta)$ are not present in the relative $a(\eta)$
making the latter a preferred form to fit. Of course, the actually
obtained fitting result is the same in both. The anisotropy will be
discussed in more detail further down. It is also of interest to
note that the SAID phase shift analysis \cite{SAID}, which does not
include the data from Ref. \cite{Betigeri01a}, is significantly
different.

In the comparison (Fig. \ref{fig:Comp_850} and Table
\ref{tab:fit_uncertainties}) we have not applied the corrections
discussed above, because of the differences between the two data
sets. It is evident that the precision and quality of cross sections
in proton-proton induced reactions is far superior to those from
neutron-proton induced reactions, where high quality data are still
lacking. Therefore, in the following considerations we do not use
individual data sets but rely rather on more global
parameterisations of almost all available data as functions of the
pion angle and momentum.

\begin{figure}
\begin{center}
\includegraphics[width=8 cm]{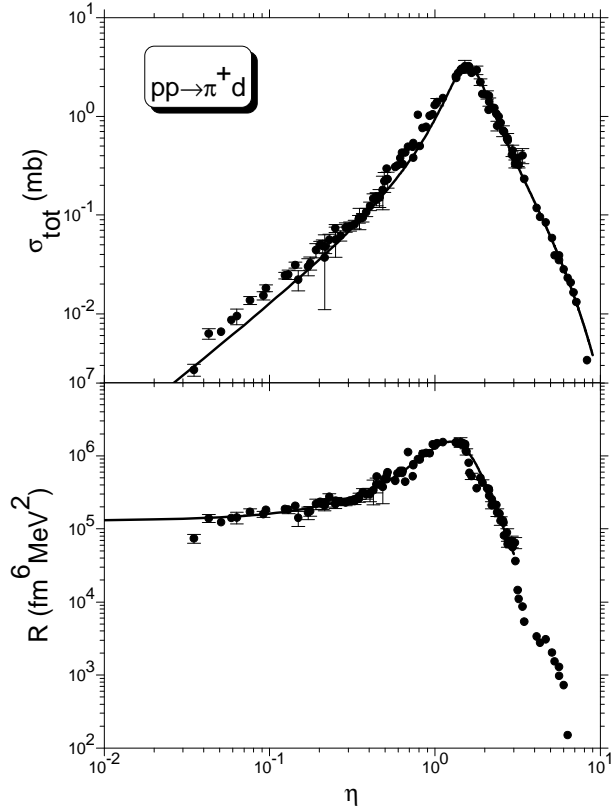}
\caption{Upper part: The total cross section for $pp\to \pi^+d$ as
function of the dimensionless pion centre of mass momentum. Lower
part: Same as the upper part but for the angle integrated sum of the
squared matrix elements, $R_{pp}$ (Coulomb corrected). The solid
curves are fits which are discussed in the text.}
\label{fig:sig_matrix_compare}
\end{center}
\end{figure}

We now consider the total cross section more globally as function of
$\eta$. In Fig. \ref{fig:sig_matrix_compare} we show almost the
world data set of cross sections for $pp \to\pi^+d$ (Refs.
\cite{Betigeri01a, Rose67, Ritchie91, Drochner96, Drochner98,
Heimberg96, Gogolev93, Preedom76, Axen76, Aebischer76, Dolnick70,
Shimizu82, Hoftietzer81, Richard-Serre79, Boswell82, Mayer85,
Nann79, Borkowski85, Yonnet93, Anderson74}) After a steep rise the
cross section falls down after the maximum which corresponds to the
centre of the $\Delta$ resonance. A global fit performed by Ritchie
\cite{Ritchie91a} underestimates the new near threshold data.
Therefore we have performed a new fit to the data body after
removing the phase space and the Coulomb effects. To get a
reasonable fit to the partially contradictory data it was found
necessary to make some choices concerning their inclusion. The
following data were omitted in the fit: The data from Ref.
\cite{Pasyuk97} (see the discussion below). Two points from Ref.
\cite{Gogolev93}, one point from Ref. \cite{Ritchie81} in the
interval $0.661\le \eta\le 0.735$, and the lowest point from Ref.
\cite{Heimberg96}.

We find the function
\begin{equation}\label{equ:Lorentz}
\sigma = \frac{a\eta^b}{\left(c-e^{d\eta}\right)^2+e}
\end{equation}
to account for the raw cross section data. This fit is included in
the upper panel of Fig. \ref{fig:sig_matrix_compare} and the fit
parametres are given in Table \ref{tab:Total_fit}.
\begin{table}[tb]
\caption{Fitted parameters of Eq. (\ref{equ:Lorentz}) to the total
cross sections.} \label{tab:Total_fit}
\begin{equation}\nonumber
\begin{array}{ccccc}
\hline
 a ({\rm mb}) & b & c & d & e \\
\hline 0.17\pm 0.03 & 1.34\pm 0.06 & 1.77\pm 0.04 & 0.38\pm 0.02 &
0.096\pm 0.02 \\ \hline
\end{array}
\end{equation}
\end{table}

Also shown in Fig. \ref{fig:sig_matrix_compare} are the Coulomb
corrected angle integrated matrix elements. These can be excellently
fitted by the squared Lorentz function
\begin{equation}\label{equ:Lorentz_squared}
R_{pp} = \left[\frac{b_1b_3^2}{(b_2-\eta)^2+b_3^2} \right]^2
\end{equation}
where $b_2$ sets the resonance position and $b_3$ the half-width.
The corresponding curve is included in the figure and the fit
parameters are given in Table \ref{tab:Fitted-parameters}. We also
tried adding a constant and $\eta$-dependent background, but the
quality of the fit remained essentially the same and the additional
terms were consistent with zero.

\begin{table}[tb]
\caption{Fitted parameters of Eq. (\ref{equ:Lorentz_squared}) to the
angle integrated and Coulomb corrected matrix elements. }
\label{tab:Fitted-parameters}
\begin{equation}\nonumber
\centering
\begin{array}{ccc}
\hline
 b_1\, ({\rm fm}^3{\rm MeV}) & b_2 & b_3  \\
\hline
 1252.347\pm 11.232 & 1.2503\pm 0.0064
 & 0.7922\pm 0.0108  \\
\hline
\end{array}
\end{equation}
\end{table}

\begin{figure}[tb]
\begin{center}
\includegraphics[width=8cm,angle=0]{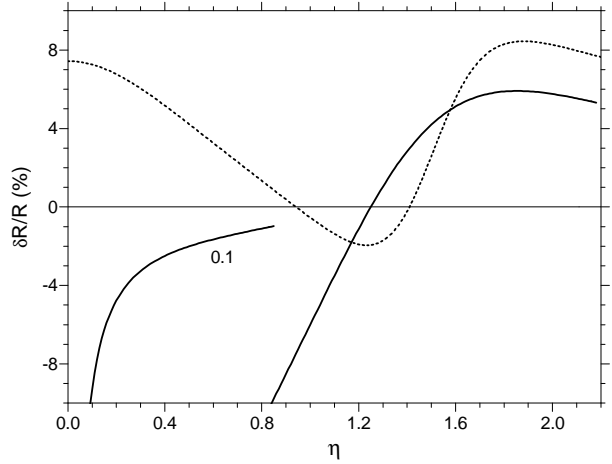}
\caption{Predictions for the relative difference of the total cross
sections of the reactions (\ref{equ:np}) and (\ref{equ:pp}). The
dashed curve is the model prediction of Ref. (\cite{Niskanen97}) for
$\delta R /R_{\rm av}$, while the solid curve is based on Eq.
(\ref{equ:prediction}) with $R_{pp}\,$.} \label{fig:prediction}
\end{center}
\end{figure}

Although the above form (especially with background terms) may look
superficially like a theoretically motivated squared amplitude, this
is not in reality the case. Since different spin amplitudes do not
mix in the cross section, odd pion waves (starting with the $p$
waves) should have a separate squared amplitude with overall
threshold behaviour $\eta^2$ (and the $\Delta(1232)$ resonance) and
even waves their own with a constant threshold behaviour (to which
even powers of $\eta$ would be added at the amplitude level). The
above form is only justified as an attempt to fit the resonant
structure with a simple function and it is, indeed, remarkable that
nearly all total cross section data can be fitted with just three
parameters. One may note that this form is not even analytically of
the correct form, which should be a function of $\eta^2$. With
functions depending on only this we were not able to get fits of
comparable quality even by increasing the number of parameters. The
data appear to require a linear term in $\eta$ in particular in the
neighbourhood of $\eta\approx 1$. Also theoretical treatments have
had trouble in giving enough cross section in this region without
destroying agreement elsewhere. With the threshold behaviour given
by this form (i.e. $dR/d\eta \not\rightarrow 0$ as $\eta \rightarrow
0$) the prediction (\ref{equ:prediction}) becomes excessively large
as can be seen from Fig. \ref{fig:prediction}.

In Eq. (\ref{equ:prediction}) the above functional form is used to
make a prediction for the difference between the reactions
(\ref{equ:np}) and (\ref{equ:pp}). This is shown together with the
model prediction of Ref. \cite{Niskanen97} in Fig.
\ref{equ:prediction}. The agreement between the two is quite
reasonable at higher energies. Considering the discussion of the
weak model dependence in that reference the difference is
significant. However, it cannot be attributed to charge dependent
interactions (both are independent of those). As discussed in the
previous section, the present prediction is based on the relation
between the initial energy and final pion momentum (albeit with
different thresholds) employing the known $pp$ data. Possible
difference in the initial state at different energies does not come
automatically except through the final momentum dependence. This
difference is most pronounced for small values of $\eta$. Even with
the correct threshold dependence the uniformly increasing matrix
elements should predict smaller $np$ than $pp$ cross sections.
However, the converse is seen in the model prediction (dashed
curve). Apparently, {\it if} one could have a varying initial
nucleon energy with a constant final momentum, the matrix elements
would decrease with the energy. This trend is overcome only in the
region of the sharp rise of the $\Delta$-resonance effect, where the
$np$ reaction lags behind and the cross section becomes smaller than
that of the $pp$ initiated reaction.

\begin{figure}
\begin{center}
\includegraphics[width=10 cm]{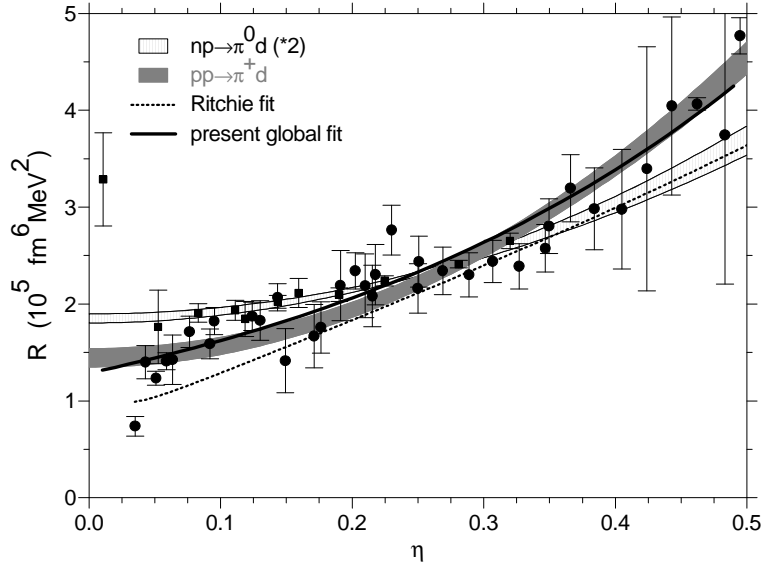}
\caption{Fit of the low energy function (\ref{equ:exfu}) to the sum
of squared matrix elements extracted from the data. The error band
represents the confidence interval on the 95$\%$ level. Data for
reaction $np\to \pi^0d$, corrected by the isospin factor of 2, are
shown as squares, those for reaction $pp\to \pi^+d$ as dots. Also
shown is the result for the fit from Ref. \cite{Ritchie91a}.}
\label{fig:matrix_low_energy}
\end{center}
\end{figure}

We now compare the angle integrated matrix elements as function of
$\eta$ in the threshold region where there are rather precise data
for both reactions. Here the cross sections for $np \to\pi^0d$ are
taken from Ref. \cite{Hutcheon90}. They are in the threshold range
up to $\eta=0.32$. In this range one should be able to apply the low
energy expansion
\begin{equation}\label{equ:exfu}
R=\alpha_0(1+\alpha_1\eta^2),
\end{equation}
where the first term is predominantly $s$-wave and the second
$p$-wave \cite{Machner98a}. The Coulomb corrected cross sections for
$pp \to\pi^+d$ are from Refs.
\cite{Betigeri01a,Rose67,Ritchie91,Drochner96,Drochner98,Heimberg96}.
Here we have restricted the range to $\eta \le 0.5$ to be
well-matched. It is worth mentioning that the cross sections from
Refs. \cite{Heimberg96} and \cite{Drochner98} agree nicely with each
other when the same Coulomb correction is applied. The Coulomb
correction used in Ref. \cite{Heimberg96} is somewhat too large (see
the discussion in Ref. \cite{Niskanen97}).  Here we have started
with the uncorrected cross sections from Ref. \cite{Heimberg97} and
applied the corrections discussed above. The deduced matrix elements
are shown in Fig. \ref{fig:matrix_low_energy} and the corresponding
fit parameters in Table \ref{tab:Matrix_low_energy}. Those for $np
\to\pi^0d$ reactions are multiplied by the isospin factor of two.
Fits with Eq. (\ref{equ:exfu}) are shown as error bands.

The excitation functions for the two reactions clearly differ from
each other with respect to magnitude and shape. Applying the same
corrections as for the data we have also extracted the matrix
elements from an earlier fit \cite{Ritchie91a} to data, which
appeared before the advent of new data in the threshold region
\cite{Drochner96, Drochner98, Heimberg96}. This is also shown in the
figure, where the different slope to the data shows the importance
of the new threshold data. Following Feynman's last point
rule\footnote{According to Feynman \cite{Feynman86} "there is a
principle that a point on the edge of the range of the data--the
last point---isn't very good, because if it was, they'd have another
point further along".} in fitting Eq. (\ref{equ:exfu}) to the data
we have excluded one point from Ref. \cite{Rose67} which is far from
the band of the other data. The same is true for the points with
smallest $\eta$ from Refs. \cite{Heimberg96} and \cite{Hutcheon90}
(see Fig. \ref{fig:matrix_low_energy}).

Differing from other $np \to\pi^0d$ measurements, the normalisation
performed by Hutcheon et al. \cite{Hutcheon90} is to the
simultaneously measured $np\to pn$ scattering cross sections. The
absolute value of this reaction was taken from the SAID phase shift
analysis \cite{SAID_NN}. The other measurements normalized their
counting rate to $pp \to\pi^+d$ cross sections. However, even the
$np\to pn$ data might not be completely independent of an assumed
isospin symmetry (see Ref. \cite{Blomgren99} and Ref.
\cite{Drochner99}). If that would be so we would expect two bands
running in parallel with only different normalisations.
Unfortunately at the present time we are not aware of such $np$
reaction data which do not suffer from normalisation problems to
such an extent as to make comparisons meaningless. 

\begin{table}[h]
\caption{Parameters from fits of Eq. (\ref{equ:exfu}) to the matrix
elements in the close to threshold
area.}\label{tab:Matrix_low_energy} \centering
\begin{tabular}{l|ll}
\hline
reaction & $\alpha_0\; ({\rm fm}^6{\rm MeV}^2)$ & $\alpha_1$ \\
\hline
$np\to \pi^0d$ & $(185\pm 2)\times 10^3$ & $3.97\pm 0.22$ \\
$pp\to\pi^+d$ & $(144.5\pm 4.7)\times 10^3$ & $8.57\pm 0.52$ \\
\hline
\end{tabular}
\end{table}

\begin{figure}[h]
\begin{center}
\includegraphics[width=8 cm,angle=0]{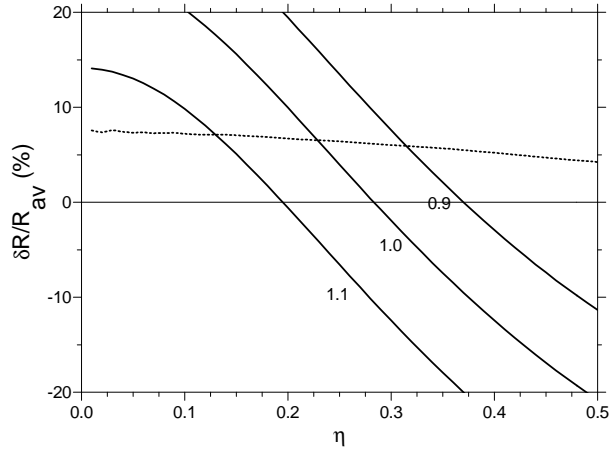}
\caption{Comparison of the relative change between the model results
of Ref. \cite{Niskanen97} (dashed curve) and the fit of Eq.
\ref{equ:exfu} (solid curves) directly and with $\alpha_0(np)$
renormalised by factors 0.9 and 1.1. } \label{fig:low_prediction}
\end{center}
\end{figure}

In spite of the above normalisation problems it should be noted that
at threshold $R_{np} > R_{pp}$ as predicted by Ref.
\cite{Niskanen97} (the dashed curve in Fig. \ref{fig:prediction}).
However, the experimental difference is larger and goes through zero
much faster than the prediction. This difference of the only
comparable data shown in Fig. \ref{fig:matrix_low_energy} can be
further elaborated into the form of the prediction of Ref.
(\cite{Niskanen97} in Fig. \ref{fig:low_prediction}. Quite clearly
the disagreement between experiment and theory cannot be removed
even by renormalising the $np$ cross sections. In the figure this
has been done by moderate factors 0.9 and 1.1. This discrepancy
might possibly be an indication of an effect of charge dependent
forces and apparently deserves a more dedicated study.

To avoid problems with the normalisation we now study the relative
anisotropy $a(\eta)$, instead.
 The nice feature of studying this quantity is that
the phase space correction and the isospin factor cancel exactly. As
seen below, to a good approximation this is true for the Coulomb
correction as well.
 However, before discussing the data we should
first have a more detailed look into the meaning of this observable.

Due to parity conservation always
\begin{equation}\label{equ:J_Breaking}
L(np)=l_\pi\pm 1 \, ,
\end{equation}
so that even isospin conserving pion waves are associated with
initial odd-triplets and {\it vice versa} odd pion waves with
even-singlet states, while for isospin breaking (\emph{CSB})
amplitudes the converse is valid\footnote{Ref. \cite{Bartlett70}
quotes additionally a superfluous and incorrect condition
$J=L(np)\pm 1$ for initial triplets, which would prevent e.g. pion
$s$-wave production.}. With the interference of opposite parities
the contribution from \emph{CSB} isospin mixing is odd in
$\cos\theta$ and does not appear in the expansion with even
polynomials, if both forward and backward angular ranges are
represented with equal weights. Charge dependence due to isotensor
forces does not mix these parities, so for this study it is
sufficient and consistent to consider only angular dependencies
symmetric about $90^\circ$.

The differential cross section can be expanded in terms of the
lowest partial waves $a_J$ (here up to the $p$-waves) as
\begin{equation}\label{equ:partial_waves}
4\pi\frac{d\sigma(\eta,\cos\theta)}{d\Omega}=\frac{1}{4} \left[
\left( |a_0|^2+|a_1|^2+|a_2|^2 \right) +\left( |a_2|^2-2\sqrt{2} \Re
\left( a_0a_2^*\right)\right)P_2\left(\cos\theta\right)\right].
\end{equation}
By comparison with Eq. (\ref{equ:Leg_expansion}) one identifies now
$1/4\left( |a_0|^2+|a_1|^2+|a_2|^2 \right)$ as the total cross
section and $1/4\left( |a_2|^2-2\sqrt{2} \Re \left(
a_0a_2^*\right)\right)$ as $A_2$. Here the $^1S_0 \rightarrow p$
amplitude $a_0$ is significantly smaller than the isotropic $s$-wave
$a_1$ or the dominant $p$-wave $^1D_2 \rightarrow p$ enhanced by the
$\Delta$ excitation with increasing energy \cite{Bugg88}. In the
case of a vanishing amplitude $a_0$, $A_2(\eta)$ would be
exclusively  of a single $p$-wave without phase shift dependent
interference. Also for a nonvanishing $a_0$, if the same Coulomb
phase is associated with the both $p$-wave amplitudes $a_0$ and
$a_2$, it will cancel off at this level of expansion and the result
would be formally the same for both reactions.

In order to be more sensitive to the $p$-wave we have to extend the
momentum range. However, then also higher Legendre-polynomials and
higher partial waves have to be considered in fitting the angular
distributions and possible interference effects involving different
Coulomb phases can arise, first the $d$-waves with the $s$-wave
pions. Their partial wave dependencies will not be discussed here.
For the general relation between the Legendre-coefficients and
higher partial wave amplitudes (also including many spin
observables) see Ref. \cite{Blankleider85}. Again both reactions
should be well matched in $\eta$. In the upper end of energies lack
of data from $np\to \pi^0d$ above $\eta=2$ limits the range of
comparison with $pp\to \pi^+d$.
\begin{figure}
\begin{center}
\includegraphics[width=8 cm]{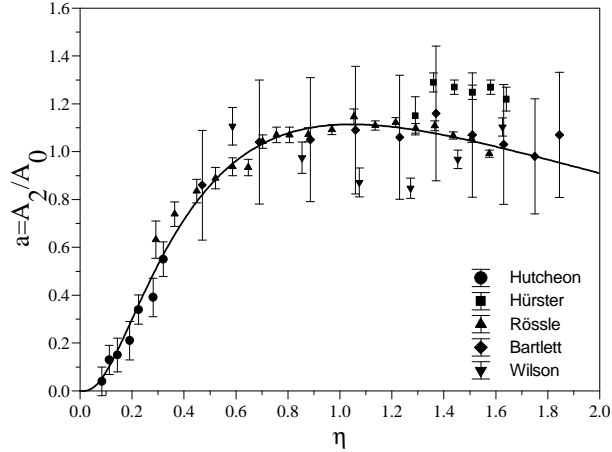}
\caption{The ratio $a=A_2/A_0$ for the reaction $np\to \pi^0d$ as
function of $\eta=p_\pi/m_\pi$. The solid curve is a fit as
discussed in the text.} \label{fig:a1_a0_np}
\end{center}
\end{figure}

The data for reaction (\ref{equ:pp}) are from Refs.
\cite{Drochner96, Drochner98, Heimberg96, Betigeri01a, Aebischer76,
Hoftietzer81, Guelmez93, Boswell82, Richard-Serre79, Ritchie91}. The
data from \cite{Dolnick70} and \cite{Norem71} were excluded, since
they are far above the other data. The data from the time reversed
reaction from \cite{Pasyuk97} were also excluded from the fit
because of the large spread within this data set (see the discussion
in Ref. \cite{Betigeri01a}). In the case of reaction (\ref{equ:np})
only the data from Refs. \cite{Bartlett70, Roessle81, Hutcheon90}
were taken into account. Those from   \cite{Wilson71a} and
\cite{Huerster80} were excluded on similar grounds as in the case of
the charged channel. Although the error bars are small (see also
Table \ref{tab:fit_uncertainties}) there seem to be systematical
uncertainties. This is demonstrated in Fig. \ref{fig:a1_a0_np}.

\begin{figure}\centering
\includegraphics[width=8 cm]{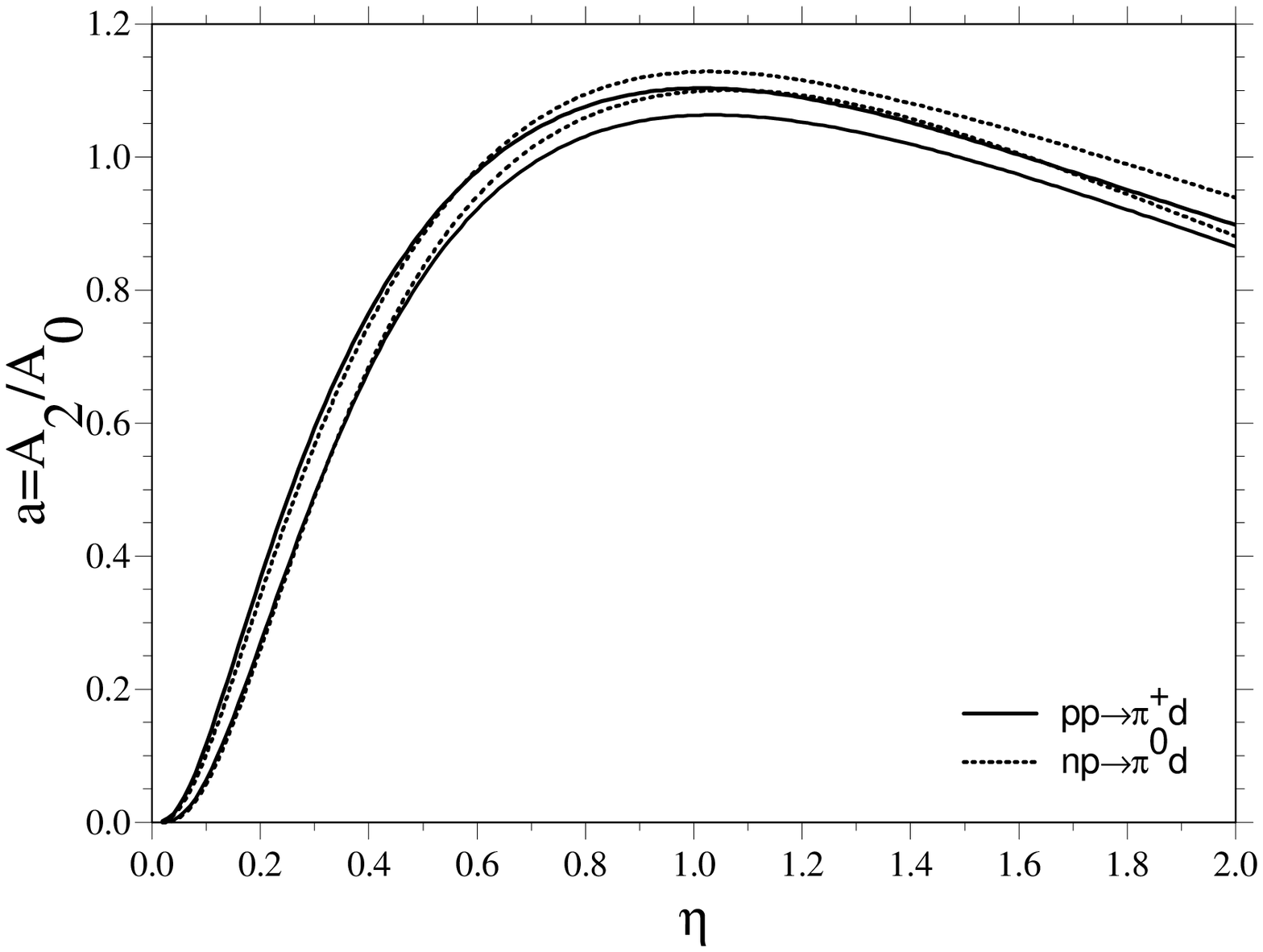}
\caption{The confidence intervals (on 95$\%$ level) of the fitted
anisotropies.} \label{fig:a2_a0_long}
\end{figure}

In order to make a comparison between the two reactions possible, we
have fitted to data a function which describes the dependence very
well, although it is not motivated by any theory. The function is
\begin{equation}\label{equ:fit_function}
a(\eta)=\alpha \exp\left\{-\frac{1}{2}\left[ \ln\left(
\frac{\eta}{\beta} \right)/\gamma\right]^2\right\}.
\end{equation}
\begin{table}
\centering \caption{Parameters obtained from fitting Eq.
(\ref{equ:fit_function}) to the anisotropies.} \label{tab:fit_param}
\begin{tabular}{llll}
\hline
reaction & $\alpha$ & $\beta$ & $\gamma$ \\
\hline
$np\to \pi^0d$ & $1.114\pm0.007$ & $1.044\pm0.017$ & $1.020\pm0.029$ \\
$pp\to\pi^+d$ & $1.111\pm 0.007$ & $1.003\pm 0.009$ & $0.926\pm 0.020$ \\
\hline
\end{tabular}
\end{table}
The fitted parameters are compiled in Table \ref{tab:fit_param}.
While the values for $\alpha$ agree with each other, the values for
$\beta$ and $\gamma$ are slightly different. This also reflects in
slightly different momentum dependencies of the anisotropy as shown
in Fig. \ref{fig:a2_a0_long}.

\begin{figure}\centering
\includegraphics[width=8 cm]{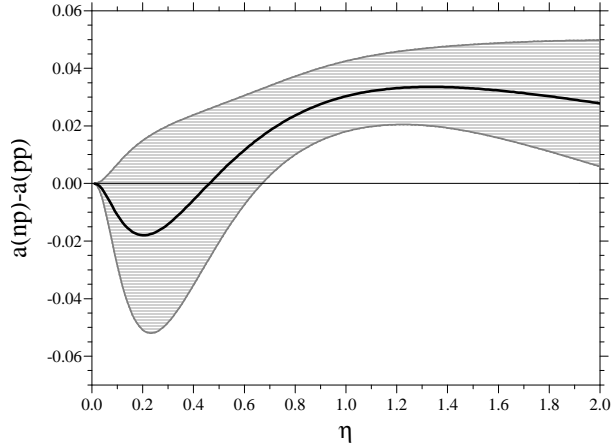}
\caption{The difference of the anisotropy for $np \to\pi^0d$ and $pp
\to\pi^+d$ (thick curve). The confidence interval (on 95$\%$ level)
is indicated by the shaded area.} \label{fig:a_pp-a_np}
\end{figure}
In order to study further the deviations of the anisotropy for the
two reactions from each other, we plot the difference of the
anisotropies in Fig. \ref{fig:a_pp-a_np}. Also the uncertainty is
shown. We can state that $a_{pp}$ is larger than $a_{np}$ in the
vicinity of the  threshold ($\eta < 0.5$) and smaller for the range
above, although the error bar is large.

\section{Discussion}
We have discussed the total and differential cross sections of the
reactions $np \to\pi^0d$ and $pp \to\pi^+d$ as functions of the pion
momentum in order to allow for a comparison with respect to isospin
symmetry. This takes into account automatically the different
thresholds but with the cost of having different initial kinetic
energies. However, this difference can be dealt with easily to first
order once Coulomb effects have been corrected for and the slight
difference in the phase space. It is found that the differential
cross section data at individual energies for the reaction $np
\to\pi^0d$ do not have the quality to make such a comparison useful.

Having first made a global fit of the total cross sections and the
associated effective matrix elements for future use, we then
compared a number of total cross sections in the threshold region.
This limitation is enforced by the lack of higher energy absolute
cross sections of the $np \to\pi^0d$ reaction. The result is that
very close to threshold $2\sigma(np \to\pi^0d)> \sigma(pp
\to\pi^+d)$, contrary to a simple "model independent" expectation
(\ref{equ:prediction}) but in agreement with a theoretical
anticipation \cite{Niskanen97}. It was not possible to get both the
size and steepness of the difference as a function of the final pion
momentum to agree with the charge independent theory.

The cross section in the interval studied is mainly due to the
transition ${^3P_1}\to {^3S_1}s$, which dominates the outgoing
$s$-wave with the partial wave amplitude $a_1$, and ${^1D_2}\to
{^3S_1}p$ in the $p$-wave amplitude $a_2$. There is also the
externally indistinguishable contribution from the deuteron $D$
state, so the shorter notation $^{2S+1}L_J\to l_\pi$ is well
justified. Apparently the threshold normalisation is dictated by the
$s$-wave amplitude (which yields the size order of the two cross
sections dependent on the initial energy), while the $p$-wave
determines the slope and thus the zero crossing point of the
difference. Simply renormalising the $np$ cross section has opposite
effects in the threshold difference and zero crossing in a
comparison of theory and experiment and is not a solution to their
discrepancy.

We then compared the relative anisotropy for both reactions, where
phase space corrections are not necessary and also the Coulomb
effects are minute. It is mainly due to the outgoing $p$-wave with
the $\Delta$ enhanced transitions ${^1D_2}\to p$.  The comparison of
the anisotropy for both reactions shows that $a(np \to\pi^0d)> a(pp
\to\pi^+d)$ although the uncertainty is large. The reason for this
is again the poor quality of the $np$ data. As an example we mention
the overlap region around $\eta\approx 0.35$ between the two data
sets \cite{Hutcheon90} and \cite{Roessle81}, which differ strongly,
as is shown in Fig. \ref{fig:a1_a0_np}. The problem of energy
resolution of a neutron beam and the absolute calibration of its
intensity may be overcome by changing the incident channel incident
channel at the cost of increasing the number of participating
nucleons. A quasi-free reaction $pd\to \pi^0dp_{s}$ with a spectator
proton $p_s$ was recently shown to be feasible in a storage ring
with a cluster target \cite{Bilger01}. Also the reaction $dp\to
\pi^0dp_{s}$ with a fast spectator proton may be considered.

\section{Acknowledgement}
This work was supported in part by the Academy of Finland (number
211592) and DAAD (Germany, number DB000379) exchange grant. We thank
C. Hanhart and K. Kilian for useful  discussions.


\end{document}